\documentclass[aps,prb,twocolumn,superscriptaddress,longbibliography, nofootinbib]{revtex4-2}
\usepackage[english,french]{babel}
\pdfoutput=1
\usepackage{graphicx,bm,amsmath,eufrak,amssymb,url}
\usepackage[usenames,dvipsnames]{color}
\usepackage[bbgreekl]{mathbbol}
\usepackage{xcolor}
\usepackage{braket}
\usepackage[colorlinks=true, breaklinks=true, linkcolor=darkblue, citecolor=darkblue, urlcolor=darkblue]{hyperref}
\usepackage{float}
\usepackage{pgf}
\usepackage[T1]{fontenc}
\usepackage{verbatim}
\usepackage{graphicx}
\usepackage[font=small,labelfont=it,justification=justified,format=plain]{caption}
\usepackage{subcaption}
\usepackage{url}
\usepackage{cancel}
\usepackage{algorithm}
\usepackage{algpseudocode}
\usepackage[normalem]{ulem}
\definecolor{darkblue}{rgb}{0, 0, 0.8}
\newcommand{\be}{\begin{equation}}
\newcommand{\ee}{\end{equation}}
\newcommand{\bea}{\begin{eqnarray}}
\newcommand{\eea}{\end{eqnarray}}
\newcommand{\ba}{\begin{align}\begin{split}}
\newcommand{\ea}{\end{split}\end{align}}

\newcommand{\am}[1]{\textcolor{blue}{#1}}

\newcommand{\change}[1]{{\color{OliveGreen} \textbf{MODIF: }#1}}

\begin{document}
\title{Hubbard physics with Rydberg atoms: using a quantum spin simulator to simulate strong fermionic correlations}
%\title{Studying the fermionic 2D-Hubbard model with neutral atoms}
\author{Antoine Michel}
\email{antoine.michel@edf.fr}
\affiliation{Electricité de France, EDF Recherche et Développement, Département Matériaux et Mécanique des Composants, Les Renardières, F-77250 Moret sur Loing, France }
\affiliation{Université Paris-Saclay, Institut d’Optique Graduate School,
CNRS, Laboratoire Charles Fabry, F-91127 Palaiseau Cedex, France}
\author{Loïc Henriet}
\affiliation{PASQAL, 7 rue Léonard de Vinci, F-91300 Massy, France}
\author{Christophe Domain}
\affiliation{Electricité de France, EDF Recherche et Développement, Département Matériaux et Mécanique des Composants, Les Renardières, F-77250 Moret sur Loing, France }
\author{Antoine Browaeys}
\affiliation{Université Paris-Saclay, Institut d’Optique Graduate School,
CNRS, Laboratoire Charles Fabry, F-91127 Palaiseau Cedex, France}
\author{Thomas Ayral}
\affiliation{Eviden Quantum Laboratory, Les Clayes-sous-Bois, France}

\selectlanguage{english}

\begin{abstract}
We propose a hybrid quantum-classical method to investigate the equilibrium physics and the dynamics of strongly correlated fermionic models with spin-based quantum processors.
Our proposal avoids the usual pitfalls of fermion-to-spin mappings thanks to a slave-spin method which allows to approximate the original Hamiltonian into a sum of self-correlated free-fermions and spin Hamiltonians. 
Taking as an example a Rydberg-based analog quantum processor to solve the interacting spin model, we avoid the challenges of variational algorithms or Trotterization methods.
\begin{comment}
\sout{on Noisy, Intermediate Scale Quantum (NISQ) processors as well as the Trotterization issues of gate-based processors.}\change{or Trotterization methods.}
\sout{At the same time, the method allows to leverage the large number of controlled particles available on such processors.}
\end{comment}
We explore the robustness of the method to experimental imperfections by applying it to the half-filled, single-orbital Hubbard model on the square lattice in and out of equilibrium.
We show, through realistic numerical simulations of current Rydberg processors, that the method yields quantitatively viable results even in the presence of imperfections: it allows to gain insights into equilibrium Mott physics as well as the dynamics under interaction quenches.
This method thus paves the way to the investigation of physical regimes---whether out-of-equilibrium, doped, or multiorbital---that are difficult to explore with classical processors.  
\end{abstract}

\maketitle

%\ab{be super rigourous in the way you call things: the platform is sometimes a processor, or a processor or computer: always use the same wording... }

%Strongly correlated electron systems display a large variety of exotic phases of matter emerging from complex quantum many-body phenomena \cite{PhysRevLett.114.176401,PhysRevLett.130.066401}.
%Numerous theoretical efforts in the past decades have led to a better understanding of some parts of the phase diagrams of these systems thanks, in particular, to many advances in numerical algorithms to investigate these systems: for instance, the single-site Hubbard model, which is believed to be a minimal model to describe high-temperature superconductors, is well characterized by numerous methods for some parameter regimes (see e.g \cite{Leblanc2015,Schafer2021} for extensive reviews of numerical results).
%Yet, some important regimes of parameters still elude any robust understanding: typically, low-temperature, doped phases of strongly-correlated models are arguably out of the reach of the most advanced methods like quantum Monte-Carlo methods \cite{Wagner2018} or embedding methods \cite{Sun2016a}.
%The situation is even worse for more realistic models with multiple orbitals---required to describe, for instance, iron-based superconductors \cite{Si2016}, or the recent Moiré superconductors \cite{Andrei2021}, let alone out-of-equilibrium phenomena like sudden quenches, which offer an additional exponential difficulty in the guise of a dynamical sign problem or a ballistic growth of entanglement.

Decades of theoretical efforts have led to tremendous progress in the understanding of the exotic phases of strongly correlated electron systems. For instance, lots is known about the physics of their minimal model, the Hubbard model \cite{Leblanc2015,Schafer2021}. Yet, the exponential difficulty of the underlying many-body problem still poses formidable challenges in low-temperature, doped phases relevant to cuprate superconductors, in multi-orbital settings relevant, for instance, to iron-based superconductors \cite{Si2016} and the recent Moiré superconductors \cite{Andrei2021}, or in out-of-equilibrium situations like sudden quenches that lead to a fast growth of entanglement \cite{Calabrese2005}.

\textit{Quantum processors}, i.e., controllable, synthetic quantum many-body systems \cite{ayral_quantum_2023}, are promising  to solve these outstanding challenges \cite{Feynman1982}.
Ultracold fermionic atoms trapped in optical lattices were already implemented more than a decade ago \cite{schneider_metallic_2008,Jordens2008,Esslinger2010,schneider_fermionic_2012,schreiber_observation_2015,Hart2015,Cheuk2016,Boll2016,mazurenko_cold-atom_2017,Tarruell2018} as the most direct, or "analog", quantum processors of fermions. They allowed to observe signatures of, for instance, Mott physics, while operating---so far---at temperatures too high to gain insights into pseudogap or superconducting phases.
In contrast, universal "digital" quantum processors rely on  quantum bits encoded on two-level or "spin-1/2" systems, and operate logic gates on them. They in principle enable the simulation of the second-quantized fermionic problems explored in  materials science \cite{Bauer2020} or chemistry \cite{Cao2019}.
% I removed \cite{gonzalez-cuadra_fermionic_2023} because it has to do with Rydberg atoms... not gate-based stuff
Yet,  early attempts are facing the physical limitations of these processors
%(dubbed Noisy, Intermediate Scale Quantum (NISQ) processors \cite{preskill_quantum_2018} \ab{I usually associate NISQ to simulator and not digital approach. Et en plus jene pense pas que ce soit pertinent de parler de NISQ ici...je virerais}) \am{Nous ne sommes pas trop d'accord, les NISQ sont à la fois analog et digital}
in terms of the number of qubits and number of gates that can be reliably executed before decoherence sets in.
Fermionic systems are particularly demanding due to the loss of locality of the Hamiltonian \cite{JordanWigner1928,Bravyi2002} or the need for auxiliary qubits \cite{Verstraete2005,Setia2018,Derby2021} that come with translating to a qubit language. 
Both constraints generically lead to longer quantum programs, and hence an increased sensitivity to imperfections.
%of NISQ processors.
To alleviate those issues, hybrid quantum-classical methods \cite{Bharti2021, Endo2021} such as the Variational Quantum Eigensolver (VQE, \cite{Peruzzo2014}) were proposed, with many developments but without clear-cut advantage so far.

%\ab{the next paragraph seems to mix many ideas. To be fixed and sharpen.}
Despite remarkable recent progress towards large-size digital quantum processors, "analog" quantum processors remain a serious alternative to explore
fermionic problems. Beyond the aforementioned ultracold atoms, 
analog platforms include systems of trapped ions and cold Rydberg atoms.
The lesser degree of control of these processors---with a fixed, specific "resource" Hamiltonian that does not necessarily match the "target" Hamiltonian of interest---is compensated for by the large number of particles that can be controlled, with now up to a few hundreds of particles \cite{scholl_quantum_2021,chen_continuous_2023, ebadi_quantum_2021}.
In addition, the parameters of the resource Hamiltonian are usually precisely controlled in time \cite{glaetzle_designing_2015, bluvstein_controlling_2021, browaeys_many-body_2020, bloch_many-body_2008,scholl_quantum_2021,gonzalez-cuadra_fermionic_2023}.
This has enabled the use of such processors to study many-body problems in several recent works 
\cite{kokail_self-verifying_2019, fermion_analog, arguello2019analogue, michel2023blueprint, bloch_many-body_2008}. %\ab{remix ideas between VQE...}
For instance, \cite{kokail_self-verifying_2019} have investigated the physics of the Schwinger model---a toy problem for lattice quantum electrodynamics---by leveraging the similarity between the symmetries of a 20-ion quantum simulator and those of the Schwinger model. 
\begin{comment}
In general, however, such a similarity between target and resource 
Hamiltonians is rare. In particular, the question of how to tackle a fermionic many-body problem with a spin-based, analog simulator is an open problem.
\end{comment}
However, finding such a similarity between target an resource Hamiltonian is rare. For instance, the question of how to tackle a fermionic many-body problem with a spin-based, analog simulator is an open problem.
%In this Letter, we propose a method to fill this gap.

%\ab{ca arrive quand meme un peu tard...;-)}
In this Letter, we propose a method to address this problem considering a specific processor, namely an analog Rydberg quantum processor \cite{henriet_quantum_2020, browaeys_many-body_2020}.
By using a self-consistent mapping between the fermionic problem and a "slave-spin" model, we circumvent the nonlocality issues related to fermion-to-spin transformations.
We show that the method allows one to compute key properties of the Hubbard model in and out of equilibrium.
We show, through realistic numerical simulations, that it does so even in the presence of hardware imperfections like decoherence, readout error and finite-sampling shot noise.

\textit{The slave-spin method.}
As a proof of concept, we consider the single-band, half-filled Fermi-Hubbard model on a square lattice. Its Hamiltonian, 
\begin{equation}
    H_\text{Hubbard} = \sum_{i,j,\sigma} t_{ij} d_{i\sigma}^{\dagger} d_{j\sigma} + \frac{U }{2} \sum_{i} (n_{i}^d - 1)^2  \ , 
      \label{Hubbard_half}
\end{equation}
contains creation (resp. annihilation) operators $d^\dagger_{i\sigma}$ (resp. $d_{i\sigma}$) that create (resp. annihilate) an electron of spin $\sigma$ on lattice site $i$, with a hopping amplitude  $t_{ij}$  between two sites (we will focus on nearest-neighbor hopping only, $t_{ij} = - t \delta_{\langle ij \rangle}$) and an on-site interaction $U$. The chemical potential was set to $\, \mu=U/2$ to enforce half-filling. 

This prototypical model of strongly-correlated electrons is hard to solve on classical processors \cite{Wu2023}, especially in out-of-equilibrium situations where the most advanced
%\ab{Ok pour cet advanced la...}
methods are usually limited to short-time dynamics. %In the following, we consider a square lattice for this model and half-filling is forced by setting $\, \mu = \frac{U}{2}$.
Instead of directly tackling this fermionic model, we thus resort to a separation of variables that singles out two degrees of freedom of the model, namely spin and charge.
This is achieved by resorting to a "slave-particle" method known as $Z_2$ slave-spin theory \cite{ruegg_mathsfz_2-slave-spin_2010}.     
We replace the fermionic operator $d^{\dagger}_{i\sigma}$ by the product of a pseudo-fermion operator $f^\dagger_{i\sigma}$ and an auxiliary spin operator $S^z_i$ ($S^{a=x,y,z}_i$  denote the Pauli spin operators), namely $d_{i\sigma}^{\dagger} = S_i^z f_{i\sigma}^{\dagger}$.
The ensuing enlargement of the Hilbert space is compensated for by imposing constraints $S_i^x+1 = 2(n^f_i-1)^2$ on each site. . In our case, namely the single-orbital, half-filled Hubbard model on a square---i.e bipartite---lattice, particle-hole symmetry holds, which is a sufficient condition for the constraint to be automatically satisfied \cite{schiro_quantum_2011}.

We then perform a mean-field decoupling of the pseudo-fermion and spin degrees of freedom $S^z_{i}S^z_{j}f_{i,\sigma}^{\dagger}f_{j,\sigma} \approx  \langle S^z_{i}S^z_{j}\rangle f_{i,\sigma}^{\dagger}f_{j,\sigma} + S^z_{i}S^z_{j} \langle f_{i,\sigma}^{\dagger}f_{j,\sigma} \rangle - \langle S^z_{i}S^z_{j} \rangle \langle f_{i,\sigma}^{\dagger}f_{j,\sigma} \rangle$. We obtain a sum of two self-consistent Hamiltonians $H \approx H_\mathrm{f} + H_\mathrm{s}$:
\begin{subequations}
\begin{alignat}{2}
    H_\mathrm{f} &= \sum_{i,j, \sigma} Q_{ij} f_{i,\sigma}^{\dagger}f_{j,\sigma}
    %+  \left(\, \mu - \frac{U}{2}\right)\sum_{i} n_{i}^f,
    \label{Ham_f}
    \\
    H_\mathrm{s} &=  \sum_{i,j} J_{ij} S^z_{i}S^z_{j} + \frac{U}{4} \sum_{i} S_i^x \label{Ham_S},
\end{alignat}
\end{subequations}
with $Q_{ij} = t_{ij} \langle S_i^z S_j^z \rangle$ and $J_{ij} = \sum_{\sigma} t_{ij} \langle f_{i,\sigma}^{\dagger}f_{j,\sigma} \rangle$.
In our particle-hole symmetric model, the constraint will be automatically fulfilled at the mean-field level \cite{yang_benchmarking_2019,de_Medici_modeling_2017}.

\begin{figure}
    \centering
    \includegraphics[width =  1 \linewidth]{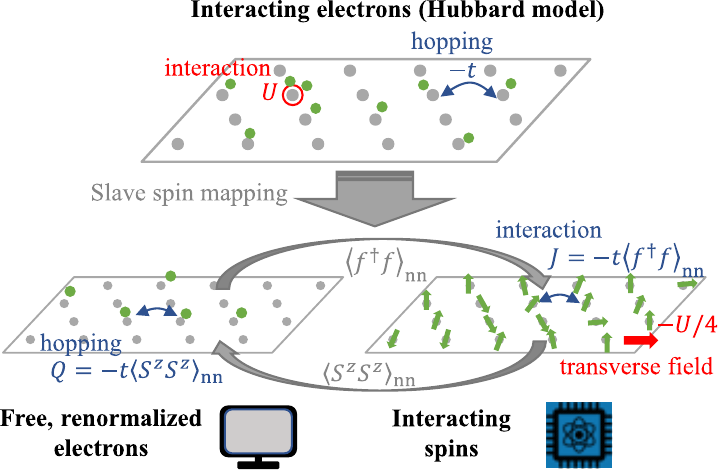}
    \caption{\textit{Slave spin mapping.} The Hubbard Hamiltonian (top) is mapped onto two self-consistently determined simpler problems: an efficiently solvable free fermionic Hamiltonian with a renormalized hopping (bottom left, $H_\mathrm{f}(Q)$ in the text), and a transverse field Ising Hamiltonian (bottom right, $H_\mathrm{s}(J)$ in the text), which we solve using a  Rydberg-based quantum processor.}
    \label{fig:loop_SSMF_final}
\end{figure}

Solving the Hubbard model using slave spin theory amounts to solving these two self-consistently defined Hamiltonians. This is done iteratively, as illustrated in Fig.~\ref{fig:loop_SSMF_final}. We start from an initial guess for the renormalized hopping $Q$ to initiate the self-consistent computation. First, we  calculate on a {\it classical} processor
the correlation function $\langle f_{i,\sigma}^{\dagger}f_{j,\sigma} \rangle$ of the pseudo-fermion problem, needed to define the spin interaction $J_{ij}$. The full derivation can be found in the Supplemental Material  section II. A.
Second, as  the spin problem is hard to solve on a classical processor, we use the analog quantum processor to compute its spin-spin correlation function.
Since $H_\mathrm{s}$ is of infinite size, we first reduce it to a finite-size problem by using a cluster mean-field approximation, as done in e.g. \cite{hassan_slave_2010} 
we solve:
\begin{equation}\label{Ham_C}
    H_\mathrm{s}^{\mathcal{C}} = \sum_{i,j \in \mathcal{C}} J_{ij} S^z_{i}S^z_{j} + \frac{U}{4} \sum_{i\in \mathcal{C}} S_i^x + \sum_{i \in \mathcal{C}} h_i S_i^z,  
\end{equation}
where $\mathcal{C}$ denotes the set of $N$ cluster sites and $h_i  = 2 z_i \overline{J} \overline{m}$ is the self-consistent mean field that mimics the influence of the infinite lattice.
Here, $z_i$ is the number of neighbors of site $i$ outside the cluster, $\overline{J}=\frac{1}{N_p} \sum_{\langle i,j \rangle \in \mathcal{C}} J_{i,j}$ is the average nearest-neighbor coupling ($N_p$ is the number of nearest-neighbor links inside the cluster) and  $\overline{m} = \frac{1}{N}\sum_{i\in\mathcal{C}} \langle S^z_i \rangle$ is the average magnetization.
This model needs to be solved iteratively by starting from a guess for the mean field $\overline{m}$.
For a given value of this mean field, the finite spin problem defined by $H_\mathrm{s}^{\mathcal{C}}$ is solved using a quantum algorithm (described below). 
This yields the correlation function $\langle S^z_i S^z_j \rangle$ and closes the self-consistent loop, which runs until convergence.
At convergence, we extract relevant observables of the original Hubbard model. For instance, the quasiparticle weight $Z$ of the original model, which measures the quantum coherence of the fermionic excitations, is obtained via the spin model's magnetization: $Z =  \overline{m}^2$ (we also have access to site-resolved magnetizations $\langle S_i^z \rangle$ and hence site-resolved quasiparticle weights).
%\change{Indeed, the slave-spin mapping implies that the original model Green's function can be rewritten as $G_{i,j}^d(\tau) = G_{ij}(t)^f G_{ij}^{S}(t)$ within the mean-field formalism. After the Fourier transform, and considering long-range interactions for the slave-spins in the metallic phase, the original Green's function is now equal to \cite{de_Medici_modeling_2017}:
%\begin{equation}
%    G_{\mathbf{k}}^d(\omega) = Z G_{\mathbf{k}}^f(\omega) + \int \frac{d\omega'}{2\pi}\frac{d \mathbf{k'}}{8 \pi^3} G_{\mathbf{k'}}^f G_{\mathbf{k}-\mathbf{k'}}^{S}(\omega - \omega')
%\end{equation}
%where $Z = \langle S_i^z \rangle \langle S_j^z \rangle = \overline{m}^2$ in the CMF approach. Therefore, $Z$ plays the role of a quasiparticle weight in this paradigm. \ab{Well... i guess i need to trust you here...;-)}}

\begin{figure}
    \centering
    \includegraphics[width =  1 \linewidth]{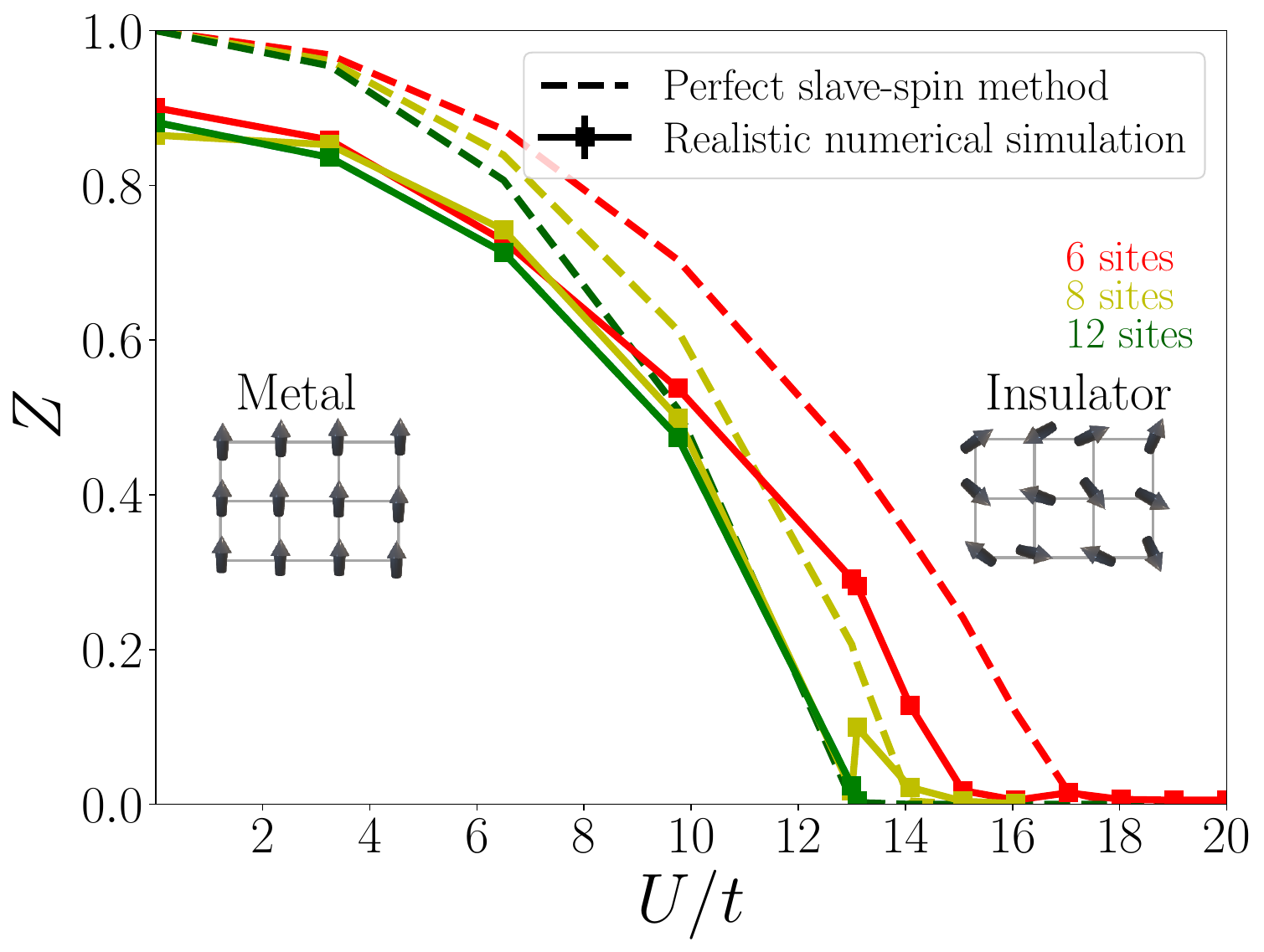}
    \caption{\textit{Mott transition observed with the slave-spin method on a realistic numerical simulation of Rydberg atoms processor.} The characteristics of the processor considered are  $\tau_\text{max} = 4$ $\, \mu$s, $\gamma = 0.02$ MHz, $N_\mathrm{s} = 150$, $\epsilon = \epsilon' = 3$\% and $5 \times 5$ loops allowed (see Supp. Mat. sections III.C). The error bar is the standard error stemming from $\epsilon$ and $N_\mathrm{s}$.}
    \label{fig:particle_weights}
\end{figure}

\textit{Quantum algorithm for the spin Hamiltonian.} Let us turn to the solution of the (cluster) spin problem, $H_\mathrm{s}^{\mathcal{C}}$.
It is nothing but the transverse-field Ising model, which could be a candidate problem for reaching quantum advantage using gate-based quantum processors \cite{Kim2023}.
As it turns out, its form is similar to the Hamiltonian realized experimentally by Rydberg atoms trapped with optical tweezers, for which the geometry of the array can be chosen at will~\cite{browaeys_many-body_2020}:
\begin{equation}\label{eq:reshamglob}
\hat{H}_\mathrm{Rydberg}= \sum_{i\ne j}\frac{C_6}{|\textbf{r}_i-\textbf{r}_j|^{6}} \hat{n}_i \hat{n}_j +  \frac{\hbar\Omega (\tau)}{2}\sum_{i} \hat{S}_i^x - \hbar \delta(\tau)\sum_{i} \hat{n}_i,
\end{equation}
where $\Omega(\tau)$ and $\delta(\tau)$ are the time-dependent Rabi frequency and laser detuning, and $C_6$ the magnitude of the interatomic van der Waals interactions; $\hat{n}_i = (I_i + S^z_i)/2$.
%Despite having the method to reach the value of $Z$ with the slave-spin theory, it is not straightforward to solve $H_\mathrm{S}(J)$ with a Rydberg atoms device. First, it is important to note that the sign of $J$ in our simulations is negative (due to the sign of t) and therefore we are dealing with a ferromagnetic transverse-field Ising Hamiltonian whereas the Rydberg Hamiltonian has an antiferromagnetic interaction 
Therefore, we can make use of the Rydberg processor to attain the ground state of $H_\mathrm{s}^{\mathcal{C}}$ using an annealing procedure \footnote{The main difference between $\hat{H}_\mathrm{Rydberg}$ and $H_\mathrm{s}^{\mathcal{C}}$ is the sign of the interaction: it is positive for Rydberg atoms, negative (since usually $t_{ij}<0$) for the slave spin problem. Thus, in practice, parameters are tuned such that the annealing procedure is performed from an initial Hamiltonian of which the system's initial state is its most excited state to the final Hamiltonian $-H_\mathrm{s}^{\mathcal{C}}$. The adiabatic theorem can also be applied for the most excited state and therefore the procedure should bring the system to (approximately) the most excited state of $-H_\mathrm{s}^{\mathcal{C}}$ \textit{i.e.} the ground state of $H_\mathrm{s}^{\mathcal{C}}$.}: we start from drive parameters $\Omega(\tau=0) = 0$ and a large positive $\delta(\tau=0)$ so that the system's native initial state $\ket{\psi_\mathrm{start}}=\ket{g}^{\otimes N}$ is the ground state of the initial Hamiltonian. We then, for a long enough annealing time, ramp the Rabi frequency and detuning to reach the final values $\hbar\Omega(\tau_\mathrm{max}) = \frac{U}{2}$, 
$\hbar\delta_i(\tau_\mathrm{max}) = \sum_{j \neq i} \frac{C_6}{r_{i,j}^6} -4\overline{J}\overline{m}z_i$. 
Optimizing the atom positions in such a way that $\frac{C_6}{r_{i,j}^6} \approx 4J_{i,j}$ (details about this optimization are in the Supplemental Material  section III. A), the final Hamiltonian will be $H_\mathrm{s}^\mathcal{C}$.
%\begin{equation}\label{Ham_target}
%H_\mathrm{target} = \sum_{i,j \in \mathcal{C},\sigma} (-J_{i,j}) S^z_{i}S^z_{j}  %-\frac{U}{4} \sum_{i\in \mathcal{C}} S_i^x- \sum_{i \in \mathcal{C}} h_i S_i^z,  
%\end{equation}
Hence, following the adiabatic theorem, the procedure should (approximately) bring the system to the ground state of $H_\mathrm{s}^{\mathcal{C}}$. We can finally measure the spin-spin correlation function on this state.

\textit{Results at equilibrium}. We implemented this self-consistent procedure with a realistic numerical simulation of a Rydberg atom processor. We repeated the computation for several values of the local Hubbard interaction $U$ to obtain the evolution of the quasiparticle weight  $Z$ as a function of $U$, as shown in Fig.~\ref{fig:particle_weights}, for cluster sizes, and hence number of atoms, of 4, 6, 8 and 12.
Besides the shot noise, intrinsic to any processor due to the measurement process (the number of measurement is noted $N_\mathrm{s}$), the main experimental limitations were considered in order to account for the true potential of current processors: dephasing noise (with a strength $\gamma$), measurement error (characterize by a percentage $\epsilon$), global detuning, finite annealing times $\tau_\mathrm{max}$ and imperfect positioning of the atoms to reproduce the right magnetic coupling (see Supplemental Material  section III. C for more details).
Despite these limitations, leading to few points being far from the noiseless result due to error accumulation, the quasiparticle weight we obtain (solid lines) is in fair agreement with the one obtained from a noiseless solution without shot noise of the spin model (dashed lines). The Rydberg processor can thus be used to get a reasonable estimate of the Mott transition, i.e the value $U_c$ when $Z$ vanishes and the systems turns Mott insulating.
While for the half-filled, single-band model studied in this proof-of-concept example, classical methods can be implemented to efficiently solve the spin model (see e.g \cite{schuler_universal_2016}), other regimes are less readily amenable to a controlled classical computation: doped regimes, multi-orbital models, and dynamical regimes. We investigate the latter regime in the next paragraph.

\begin{figure}
    \centering
    \includegraphics[width = 1. \linewidth]{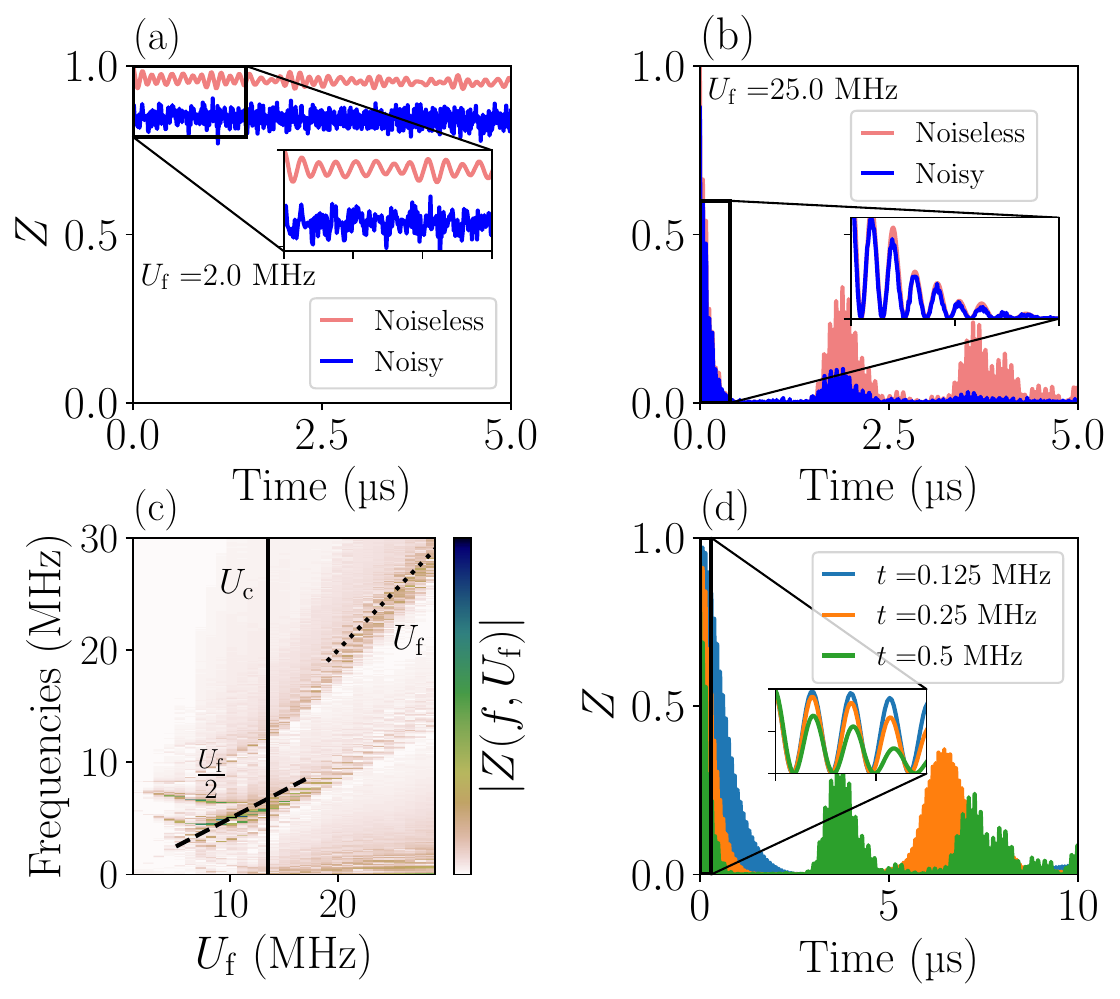}
    \caption{\textit{Dynamical response of the quasiparticle weight after an interaction quench.} $N=12$ spin cluster. Time evolution of $Z$ for (a) $U_\mathrm{f} = 2$ MHz and (b) $U_\mathrm{f} = 25$ MHz. The red line shows the noiseless annealing solution and the blue line a realistic numerical simulation on Rydberg atoms processor ($\gamma = 0.02$ MHz, $\epsilon=\epsilon' = 3\%$, $N_\mathrm{s} = 150$ shots, realistic Ising interactions and a global detuning are imposed).
    (c) Fourier transform amplitude $|Z(f,U_\mathrm{f})|$. 
    %The main frequency of the response follows these two curves.
     The vertical black line shows the equilibrium critical value $U_c$ as computed from Fig.~\ref{fig:particle_weights}.
     (d) Impact of the hopping terms $t$ on the damping of the response of $Z$ after a quench ($U_\mathrm{f} = 13 \approx U_c$). The blue, orange and green lines represent the result for $t = 0.125$, $0.25$ and $0.5$ MHz, respectively.  }
    \label{fig:quench}
\end{figure}

\textit{Results out of equilibrium.} 
We thus turn to a dynamical setting to emphasize the potential advantage brought by the use of quantum processors in this slave-spin framework.
Starting from a non-interacting ground state ($U=0$ \am{MHz}), we suddenly switch on the value of the local interaction to a final value $U_\mathrm{f}$.
Our goal is to validate that the method implemented on a realistic quantum processor recovers the phenomenology observed in previous experimental and theoretical studies of quenched Hubbard systems \cite{greiner_collapse_2002, kollath_quench_2007, schachenmayer_atomic_2011,lacki_dynamical_2019,
eckstein_thermalization_2009,iyer_coherent_2014,
schiro_quantum_2011,will_observation_2015,
riegger_interaction_2015, yang_benchmarking_2019}, such as the collapse and revival oscillations of various observables in the $U_\mathrm{f} \gg U_c$ regime, with a $2\pi/U_\mathrm{f}$ period, and a damping that increases with bandwidth. In the $U_\mathrm{f} \ll U_c$ regime, overdamped oscillations have been observed (see \cite{eckstein_thermalization_2009} for instance). 

Here, we look for this phenomenology in the time evolution of the quasiparticle weight $Z$.
Within the slave-spin method applied to the single-site Hubbard model at half-filling, interaction quenches are simple to implement: translation invariance on the lattice makes the dynamics of pseudo-fermions trivial when starting from an eigenstate (see \cite{schiro_quantum_2011} and the Supplemental Material  section I. D). 
Thus, only the dynamics of the spin model are of interest: the procedure boils down to quenching the value of the transverse field in Eq.~\eqref{Ham_C} from $0$ to $U_\mathrm{f}/4$.
On the Rydberg processor considered here, this means switching the Rabi frequency from zero up to the desired value to obtain $U_\mathrm{f}$.
In practice however, the switch-on time is not instantaneous due to the finite temporal response of the optical modulators (about $50$ ns to switch from $0$ MHz to $U_\mathrm{f} = 2$ MHz): we include this effect in our calculations.
One then directly measures $\langle S_i^z \rangle$ for different evolution times. 
In Fig.~\ref{fig:quench}, we show the oscillations observed numerically for a cluster of $12$ sites, with and without including the noise.
The upper panels present the oscillation of $Z$ as a function of time after a quench to $U_\mathrm{f} = 13$ MHz (a) and to $U_\mathrm{f}= 25$ MHz (b).
From Fig.~\ref{fig:particle_weights}, we know that the phase transition for such a cluster is $U_c \approx 13.5$ \am{MHz}.
In the case of $U_\mathrm{f} = 25$ MHz ($U_\mathrm{f} \gg U_c$), we observe the damped oscillations, both in the noiseless and the noisy setting. 
Because of the dephasing noise present in the experiment and included in the simulation, the agreement between the noiseless and noisy curves worsens with time. However, during the first microseconds of observations, we recover a nearly perfect oscillation from which we can extract the frequency (insets in (a) and (b)).
For $U_\mathrm{f} = 13$ MHz ($U_\mathrm{f} \approx U_c$), we see that $Z$ quickly reaches a value $\approx 0.1$ (slightly higher than the $Z$ obtained for this value of $U$ at equilibrium), around which it oscillates. 
Panel (c) exhibits the Fourier transform of $Z(\tau)$ for various $U_\mathrm{f}$ for the exact slave-spin method (namely with an exact solution of the spin model).
For $U_\mathrm{f} < U_c $, components at $\omega = U_\mathrm{f}/2$ can be identified along with other contributions, while for $U_\mathrm{f}>U_c$, the component at  $\omega = U_\mathrm{f}$  dominates the spectrum.
This is expected from the physics of the Mott transition in the Hubbard model: above the transition, the single-particle spectrum displays a Mott gap of $U_\mathrm{f}$, while below it excitations between the quasiparticle band and the emerging Hubbard bands (with energy $U_\mathrm{f}/2$), and within the quasiparticle band, are possible.
Finally, panel (d) confirms the expected increase of the damping of oscillations with the hopping strength $t$.

%\textit{Conclusion.}
%\ab{teh conclusion essentially repeats evrything which is in the paper. I would shorten it and re-use some of the punchy sentences in the intro.}
In conclusion, we introduced a hybrid quantum-classical method that does not suffer from the usual overheads of translating fermionic problems to spin problems, namely long quantum evolutions (due to nonlocal spin terms) or auxiliary quantum degrees of freedom.
This is made possible by using a slave-spin mapping that turns the difficult fermionic problem into a free, and thus efficiently tractable, fermion problem that is self-consistently coupled to an interacting, yet {\it local} spin problem. 

Here, to solve the spin problem, we considered an {\it analog} Rydberg-based quantum processor that naturally implements the effective transverse-field Ising spin Hamiltonian appearing in the slave-spin approach.
Its analog character allows one to circumvent the issues associated with gate-based algorithms, like trotterization when performing time evolution, and the annealing algorithm proposed here avoids the pitfalls of today's widespread variational algorithms like VQE or its temporal counterparts.
With the large number of Rydberg atoms that can be controlled in current experiments, this original approach could help tackling problems of (cluster) sizes unreachable to classical computers, without suffering from the limitations that have been pointed out \cite{Tindall2023,Begusic2023,Kechedzhi2023,torre2023dissipative} in a recent quantum-advantage experimental claim \cite{Kim2023}:
the atoms can be placed in a 2D array that has high connectivity compared to the quasi-1D connectivity of the experiments, making tensor network approaches difficult \cite{Tindall2023}, and the correlation lengths that can be attained experimentally in a similar context (up to 7 lattice sites, \cite{scholl_quantum_2021}, i.e 49 spins) also raise the bar for approaches that rely on a smaller effective size \cite{Begusic2023,Kechedzhi2023,torre2023dissipative}.
%\change{
%\ab{lien avec la phrase d'avant et quel est l'idee: jamais introduite avant?}
%For that matter, the TFIM problem does not suffer from the quasi-1D character or short correlation lengths that have been exploited  to simulate classically a recent quantum-advantage experimental claim \cite{Kim2023}: the square lattice has a high (4) connectivity and the correlation lengths that have been measured experimentally in a similar context attain up to 7 lattice sites \cite{scholl_quantum_2021}, corresponding to a 49-atom effective cluster.
%} 
This proposal calls for further investigations.
An important step would be an experimental validation with larger atom numbers than the 12 atoms we simulated here.
Other improvements involve the slave-spin method itself: doped regimes (relevant to cuprate materials), multiorbital models \cite{demedici_orbital-selective_2005} (relevant to iron-based superconductors, where orbital-selective effects may appear \cite{PhysRevLett.112.177001}) pose various technical difficulties that warrant further theoretical developments.
In particular, the fulfillment of the constraint to ensure the states remain in the physical subspace becomes more difficult in these regimes than in the half-filled, single-band case that we studied here. Going beyond the mean-field decoupling of the pseudo-fermion and spin variables is also another interesting avenue.

\acknowledgments
We acknowledge fruitful discussions with Louis-Paul Henry, Joseph Mikael, Marco Schir\`o and Thierry Lahaye.
This work was supported by the European Union's Horizon 2020 research and innovation programs under grant  agreement No. 817482 (PASQuanS) and No. 101079862 (PASQuanS2), the European Research Council (Advanced grant No. 101018511-ATARAXIA), and the European High-Performance Computing Joint Undertaking (JU) under grant agreement No 101018180 (HPC-QS).
It was also supported by EDF R\&D, the Research and Development Division of Electricité de France under the ANRT contract N°2020/0011.
Simulations were performed on the Eviden Qaptiva platform (ex Atos Quantum Learning Machine).

%%%%%%%%%% Merge with supplemental materials %%%%%%%%%%
\clearpage
\begin{center}
\textbf{\large Supplemental material:\\ Hubbard physics with Rydberg atoms: using a quantum spin simulator to simulate strong fermionic correlations}
\end{center}

\twocolumngrid

\section{Summary of the slave-spin method}\label{Appendix}

\subsection{A reminder of the main equations}

We choose the most simple form of slave spins, introduced in Ref.~\cite{ruegg_mathsfz_2-slave-spin_2010}. We recall its main steps below.
We replace the fermionic operator $d^{\dagger}$ by the tensor product of a pseudo fermion operator (that follows the same anticommutation rules as $d^{\dagger}$) and an auxiliary spin field 
\begin{equation}
    d_{i\sigma}^{\dagger} = S_i^z f_{i\sigma}^{\dagger},
\end{equation}
where $S^z_i$ is the Pauli-$Z$ operator at site $i$ (later $S^a_i$, with $a \in \{x,y,z\}$, will denote the Pauli spin operators), and $f_{i\sigma}^\dagger$ and $f_{i\sigma}$ denote fermionic operators called pseudo-fermions.  
The $d$ and $f$ operators obey fermionic anticommutation relations due to the spin commutation relations.

By substituting, in $H_\mathrm{Hubbard}$, the original fermionic operators by new spin and pseudo fermion degrees of freedom, we effectively enlarge the Hilbert space where the new Hamiltonian, $H'_\mathrm{Hubbard}$, acts.
In practice, we want to map the original problem $H_\mathrm{Hubbard}$ to a Hilbert space of same size by looking at a restriction of the new Hamiltonian, $H'_\mathrm{Hubbard}$, on a restricted portion of the new Hilbert state, which is called the physical subspace. 
This is achieved by imposing a constraint: on each site $i$, we impose the relation \cite{ruegg_mathsfz_2-slave-spin_2010}:
\begin{equation}\label{constraint_SSMF}
    \left(n_{i\uparrow}^f + n_{i\downarrow}^f - 1 \right)^2 = \frac{S^x_i+1}{2}
\end{equation}
to hold for the "physical states".
Among the eight possible local states, only four states (i.e. the same number of original local states) verify this constraint:
\begin{subequations}\label{eq:constraint_states}
\begin{alignat}{3}
    \ket{n_{i}^d=0}&=\ket{S^x = 1, n_{i}^f = 0}, \\
    \ket{n_{i,\sigma}^d=1}&= \ket{S^x= -1, n_{i,\sigma}^f = 1},\,\sigma= \uparrow, \downarrow\\
    \ket{n_{i}^d=2} &= \ket{S^x= 1, n_{i}^f = 2}.
\end{alignat}
\end{subequations}
Assuming this constraint is satisfied in the physical subspace, we can transform the original Hubbard Hamiltonian, expressed as
\begin{align}
\begin{split}\label{Hubbard_half2}
     H_\mathrm{Hubbard} =& \sum_{i,j,\sigma} t_{i,j} d_{i\sigma}^{\dagger} d_{j\sigma} + \frac{U}{2} \sum_{i} (n_{i}^d - 1)^2 \\
     \;\;+ & \left(\, \mu - \frac{U}{2} \right)\sum_{i} n_{i}^d,
\end{split}
\end{align}
to the following transformed Hamiltonian:
\begin{align}
\begin{split}\label{Ham_SS}
    H'_\mathrm{Hubbard} =& \sum_{i,j,\sigma} t_{i,j}  S^z_{i}S^z_{j}f_{i,\sigma}^{\dagger}f_{j,\sigma} + \frac{U}{2} \sum_i \left(\frac{S^x_i + 1}{2}\right) \\
     \;\;+ & \left(\, \mu - \frac{U}{2}\right)\sum_{i} (n_{i,\uparrow}^f + n_{i,\downarrow}^f),
\end{split}
\end{align}
via substitution of equality~\eqref{constraint_SSMF} in the interaction term of (\ref{Hubbard_half2}). It is straightforward to see that $n_i^d = n_i^f$ considering (\ref{eq:constraint_states}).
At this point, no approximations have been made.

The next step is then to decouple fermions and spins with a mean-field approach
\begin{align}
    \begin{split}
    S^z_{i}S^z_{j}f_{i,\sigma}^{\dagger}f_{j,\sigma} \approx&  \langle S^z_{i}S^z_{j}\rangle f_{i,\sigma}^{\dagger}f_{j,\sigma} + S^z_{i}S^z_{j} \langle f_{i,\sigma}^{\dagger}f_{j,\sigma} \rangle -\\ &\langle S^z_{i}S^z_{j} \rangle \langle f_{i,\sigma}^{\dagger}f_{j,\sigma} \rangle.
    \end{split}
\end{align}
Therefore, Eq.~\eqref{Ham_SS} can be expressed as a sum of two Hamiltonians (neglecting constant terms and considering half-filling) $H'_\mathrm{Hubbard} = H_\mathrm{s} + H_\mathrm{f}$ with $H_\mathrm{s} = \sum_{i,j,\sigma} t_{i,j}\langle f_{i,\sigma}^{\dagger}f_{j,\sigma} \rangle S_i^z S_j^z + \frac{U}{4}\sum_i S_i^x$,  a Transverse Field Ising Model (TFIM) and $H_\mathrm{f} = \sum_{i,j,\sigma} t_{i,j} \langle S^z_{i}S^z_{j}\rangle  f_{i,\sigma}^{\dagger}f_{i,\sigma}$ describing the free renormalized elctrons.
The correlators $\langle f_{i,\sigma}^{\dagger}f_{j,\sigma} \rangle$ and $\langle 
S^z_{i}S^z_{j}\rangle$ are obtained via auto-coherent loops until convergence is reached.

\subsection{Fulfillment of the constraint}
When performing loops described above, one must ensure that the constraint Eq.~\ref{constraint_SSMF} is imposed on each site. It is trivially enforced in particle-hole symmetric cases (which includes our setting, namely the single-orbital, half-filled Hubbard model on a square---i.e bipartite---lattice) \cite{schiro_quantum_2011}. In practice, the mean-field simplification leads to 
\begin{equation}\label{eq:constraint_mean}
    \langle(n_{i,\uparrow}^f+ n_{i,\downarrow}^f - 1)^2\rangle_\mathrm{f} = \left\langle \frac{S_i^x + 1}{2}\right\rangle_\mathrm{s}.
\end{equation}

In the general case, this mean-field equality can be enforced on all sites by using a Lagrange multiplier $\lambda_i$: one adds a term $H_{\lambda} = \sum_i \lambda_i ((n_i-1)^2-\frac{S_i^x+1}{2})$ to $H_\mathrm{s} + H_\mathrm{f}$ and optimize numerically all $\lambda_i$'s to reach:
\begin{equation}
    \frac{\partial \text{ln}(Tr(e^{-\beta (H_\mathrm{s} + H_\mathrm{f} +H_{\lambda})}))}{\partial \lambda_i} = 0.
\end{equation}
In the case of translational invariance, the problem can be reduced to optimizing only one parameter $\lambda = \lambda_i$.

In our case, $\lambda$ should be zero to respect the symmetry of the energy spectrum around $0$ (consequence of the particle-hole symmetry) and the constraint is fulfilled trivially \cite{yang_benchmarking_2019, de_Medici_modeling_2017}.
%% \am{J'ai changé la ref qui n'était pas bonne}.

\subsection{Variants towards a multiorbital case}
The $Z_2$ slave spin theory used here is one among others.

Another related slave-spin approach \cite{demedici_orbital-selective_2005, hassan_slave_2010} consists in enlarging the Hilbert space with the spin operator $S_{i,\sigma}^z$ such as 
\begin{equation}
    d_{i,\sigma}^{\dagger} = f_{i,\sigma}^{\dagger}S_{i,\sigma}^z 
\end{equation}

In this method, the physical states are $\ket{n_{i,\sigma}^d = 1} \Leftrightarrow \ket{n_{i,\sigma}^f = 1, S_{i,\sigma}^z = 1}$ and $ \ket{n_{i,\sigma}^d = 0} \Leftrightarrow \ket{n_{i,\sigma}^f = 0, S_{i,\sigma}^z = -1}$.
The constraint to be satisfied to only span physical states is then $n_{i,\sigma} = S_{i,\sigma}^z + \frac{1}{2}$.

While this method lends itself quite naturally to multiorbital models (see e.g \cite{demedici_orbital-selective_2005}), the additional $\sigma$ dependency of the slave-spin operators (compared to the $Z_2$ slave spins considered in our work) leads to an effective model which is more difficult to relate to existing experimental platforms.

\subsection{Dynamics in slave-spin theory} \label{subsec:dynamics}

In this subsection, we review how the slave-spin formalism extends to the time-dependent case.

After the introduction of the slave variables, we are considering the Hamiltonian:
\begin{equation}
H'_\text{Hubbard} = \sum_{i,j,\sigma} S_i^z S_j^z f_{i,\sigma}^{\dagger} f_{j,\sigma} + \frac{U}{4} \sum_i S_i^x
\end{equation}
(Eq.~\eqref{Ham_SS} at half-filling and neglecting the constants).
At the mean-field level, the time-dependent solution of the Schrödinger equation
is of the form $ \ket{\Psi(\tau)} = \ket{\Phi_\text{f}(\tau)}\ket{\Psi_\text{s}(\tau)}$ with $\ket{\Phi_\text{f}(\tau)}$ (the time will be defined as $\tau$ to avoid confusion with the hopping) governed by a Schrödinger evolution with time-dependent Hamiltonians:
\begin{align}
    \begin{split}
H_\text{f}(\tau) &= \sum_{i,j,\sigma} t_{i,j} \langle S_i^z S_j^z \rangle (\tau)f_{i,\sigma}^{\dagger}f_{j,\sigma}\\
H_\text{s}(\tau) &= \sum_{i,j,\sigma} t_{ij} S_i^z S_j^z \langle f_{i,\sigma}^{\dagger}f_{j,\sigma} \rangle (\tau) + \frac{U}{4} \sum_i S^x_i.
\end{split}
\end{align}
The initial state is of the form $\ket{\Psi(\tau = 0)} = \ket{\Phi_\text{f}(\tau=0)}\ket{\psi_\text{s}(\tau = 0)} $ with $\ket{\Phi_\text{f}(\tau=0)}$ (respectively $\ket{\Phi_\text{f}(\tau=0)}$) the ground states of $H_\text{f}(\tau < 0)$ (respectively $H_\text{s}(\tau < 0))$ found with the mean-field slave-spin procedure.
To solve these coupled equations, we a priori need to compute correlators $\langle S_i^z S_j^z \rangle (\tau)$ and $\sum_{\sigma}\langle f_{i,\sigma}^{\dagger}f_{j,\sigma} \rangle (\tau)$ and use them to construct $H_\text{f}(\tau)$ and $H_\text{s}(\tau)$. We should then evolve the wavefunctions to obtain correlators for a time $\tau + d\tau$ and so on.

In fact, as stated in \cite{schiro_quantum_2011}, the dynamics of the pseudo-fermions is trivial if our system is translation invariant (i.e $t_{ij} = t_{i-j} $).
In this case, indeed, $\langle S_i^z S_j^z \rangle (\tau) = g_{i-j}(\tau)$. Thus, 
\begin{equation}
    H_\text{f}(\tau) = \sum_{ij,\sigma} t_{i-j}g_{i-j}(\tau) f_{i,\sigma}^{\dagger}f_{j,\sigma}.
\end{equation}
This Hamiltonian is then diagonal in the Fourier space
\begin{equation}
    H_\text{f}(\tau) = \sum_k \epsilon_k(\tau) f_k^{\dagger} f_k,
\end{equation}
with $\epsilon_k(\tau)$ the Fourier transform of $ t_{i-j}g_{i-j}(\tau)$, and $n_k = f_k^{\dagger}f_k$, with $f_k \propto \sum_i e^{i k R_i} f_i$. We denote as $\ket{\Phi_{\alpha}}$ the Fock states of the system associated with the transformed operators, $f_k$. They are the eigenstates of $H_\text{f}(\tau)$ at any time $\tau$.

The initial state $\ket{\Phi_{\alpha_0}}$ is the ground state of the system.
Let's consider the time evolution of an arbitrary state $\ket{\Phi_\text{f}(\tau)}$, we can decompose $\ket{\Phi_\text{f}(\tau)} = \sum_{\alpha} c_{\alpha}(\tau) \ket{\Phi_{\alpha}} $. Therefore,
\begin{align}
    \begin{split}
        \sum_{\alpha} i\partial_t c_{\alpha}(\tau) \ket{\Phi_{\alpha}} &= \sum_{\alpha} c_{\alpha}(\tau)H_\text{f}(\tau) \ket{\Phi_{\alpha}} \\
        &= \sum_{\alpha} c_{\alpha}(\tau)E_{\alpha}(\tau)\ket{\Phi_{\alpha}} 
    \end{split}
\end{align}
One can project onto $\bra{\Phi_{\alpha}(\tau)}$:
\begin{equation}
i \partial_t c_{\alpha}(\tau) = c_{\alpha}(\tau)E_{\alpha}(\tau).
\end{equation}
Thus, $c_{\alpha}(\tau) = c_{\alpha}(\tau = 0)e^{-i \int_0^t E_{\alpha}(\tau')d\tau'}$.

Therefore, starting from the groundstate, for $\alpha \neq \alpha_0$, $c_{\alpha}(\tau) = 0$ and $c_{\alpha_0}(\tau) = e^{-i\phi(\tau)}$. At the end of the day, 
\begin{equation}
    \ket{\Phi_\text{f}(\tau)} = e^{-i\phi(\tau)}\ket{\Phi_\text{f}(\tau = 0)}
\end{equation}

and the renormalized fermionic system remains in the groundstate up to global phase, meaning that $\sum_{\sigma}\langle f_{i,\sigma}^{\dagger}f_{j,\sigma} \rangle (\tau) = \sum_{\sigma} \langle f_{i,\sigma}^{\dagger}f_{j,\sigma} \rangle_0$ is independent of time. This enables us to only consider the correlator $\langle S_i^z S_j^z \rangle (\tau)$ during the quench.

The link between eigenenergies of $H(U_\mathrm{f})$ and the frequency of oscillations can be derived: the initial state is $\ket{\psi_\mathrm{s}(\tau < 0)}$, the groundstate of $H_\mathrm{s}(U=0)$. We can decompose it in the basis of $H(U_\mathrm{f})$ eigenstates : $\ket{\psi_\mathrm{s}(\tau < 0)} = \sum_k a_k \ket{E_k}$ where $\ket{E_k}$ are eigenstates of $H(U_\mathrm{f})$ corresponding to an eigenenergy $E_k$. Let's now consider the value of an observable $\hat{O}$ through time. We obtain:
\begin{align}
    \begin{split}
        \langle \hat{O} \rangle (\tau) &= \bra{\psi_\mathrm{s}(\tau < 0)}e^{iH(U_\mathrm{f})\tau} \hat{O} e^{-iH(U_\mathrm{f})\tau} \ket{\psi_\mathrm{s}(\tau < 0)}\\
        &= \sum_{k,k'}a_k^{*}a_{k'}\bra{E_k} e^{iH(U_\mathrm{f})\tau} \hat{O} e^{-iH(U_\mathrm{f})\tau} \ket{E_{k'}}\\
        & = \sum_{k,k'}a_k^{*}a_{k'}e^{i(E_k -E_{k'})\tau} \bra{E_k}  \hat{O}  \ket{E_{k'}}
    \end{split}
\end{align} 
Therefore, frequencies of oscillations of any observable only depend on differences between eigen energies of $H(U_\mathrm{f})$.
\subsection{Dynamics and constraint fulfillment}
In the $Z_2$ slave-spin theory, one can define the projector $Q_i = \Big ( \frac{S_i^x+1}{2}-(n_i-1)^2 \Big )^2 = \frac{1-S_i^x e^{i\pi n_i}}{2} $ such that $Q_i\ket{\Psi} = 0$ iff $\ket{\Psi}$ respects the constraint Eq.~\eqref{eq:constraint_states}.
Using the fact that $[H_\mathrm{Hubbard}',\prod_i Q_i ]=0$, the constraint is fulfilled during the quench dynamics because $\ket{\Psi (\tau = 0)}$ is chosen to impose $\prod_i Q_i\ket{\Psi(\tau = 0)}=0$ (see \cite{schiro_quantum_2011,ruegg_mathsfz_2-slave-spin_2010} for more details).

\section{Solution of the two coupled subproblems}

In this section, we show how we solve numerically Eq. (2a) and Eq. (2b) in main text to obtain matrices $J$ and $Q$. For $H_\text{s}$, we describe the embedding of the cluster mean-field theory.

\subsection{Solving the quadratic fermionic Hamiltonian $H_\mathrm{f}$ for $J$} \label{subsec:bogoliubov}

To compute $J_{ij}$, we need to compute the one-particle density matrix 
\begin{equation}
    G_{ij} = \sum_{\sigma} {}_{\mathrm{f}}\bra{\psi_0}f_{i,\sigma}^{\dagger}f_{j,\sigma}\ket{\psi_0}_{\mathrm{f}}.
\end{equation}
%$\langle f_{i,\sigma}^{\dagger}f_{j,\sigma} + f_{j,\sigma}^{\dagger}f_{i,\sigma} \rangle$.

%The ground state $\ket{\psi_0}_{\mathrm{f}}$ of Eq.~\eqref{Ham_f} can be efficiently computed as $H_\mathrm{f}$ is quadratic in the creation and annihilation operators.
$H_\mathrm{f}$ can be rewritten as a matrix product:
\begin{equation}\label{eq:Hf_matrix_form}
H_\mathrm{f} = F^{\dagger}QF
\end{equation}
with $F^{\dagger} = (f_{1,\downarrow}^{\dagger},f_{1,\uparrow}^{\dagger},  f_{2,\downarrow}^{\dagger}, \dots)$ and $Q$ a Hermitian, $N\times N$, matrix.
$Q$ can be diagonalized numerically: $Q = LDL^{\dagger}$, with $D= \mathrm{diag} \lbrace \lambda_1, \lambda_2, \dots, \lambda_N  \rbrace$.
If the number of sites is even, the trace of $D$ vanishes and we obtain as many $\lambda_i <0$ as $\lambda_i > 0$.
It leads to define $C^{\dagger} = F^{\dagger}L \iff C = L^{\dagger}F$ and a diagonal form of $H_\mathrm{f}$ is obtained
\begin{equation}
    H_\mathrm{f} = \sum_{i,\sigma}\lambda_i c_{i,\sigma}^{\dagger}c_{i,\sigma}.
\end{equation}

The ground-state energy of this Hamiltonian is the sum of negative $\lambda_i$ and the groundstate is then a Slater determinant $\ket{0101\dots01}_\mathrm{C}$ in the $c$ basis with $1$ corresponding to negative energies and $0$ otherwise. To go back in the $f$ basis, one can use the $L$ matrices:
\begin{align}
G_{ij}
        &= \bra{\psi_0}\sum_{k,k',\sigma} L^{\dagger}_{k,i}L_{j,k'}c^{\dagger}_{k,\sigma}c_{k',\sigma}\ket{\psi_0}\\
        &= \sum_{k,k',\sigma}\delta_{k,k'} L^{\dagger}_{k,i}L_{j,k'}\langle c^{\dagger}_{k,\sigma}c_{k',\sigma}\rangle\\
        &= \sum_{k,\sigma} L^{*}_{i,k} n_{k,\sigma} L^{t}_{k,j}.
\end{align}
with $n_{k,\sigma}$ equal to $1$ for $k$ indices where $\lambda_k < 0$.
Numerically, it means that only matrices $L$ and eigenvalues $\lambda_i$ are needed to compute $J$.
This part can be dealt with a classical quantum computer 
%or High Performance Computer (HPC)
as it has a polynomial complexity. Going into the thermodynamic limit makes it easier as the system is really translation invariant and the Hamiltonian is then diagonal in the Fourier space. In our work, we choose to solve $H_\mathrm{f}$ considering boundaries to show the effect of a finite-size system on the method. Further developments could be done to simplify this step and consider an infinite-size system.

%On the other hand, $H_\mathrm{S}^{(C)}$ needs to be solved by a quantum computer because of exponential complexity with the number of sites.

\subsection{Solving the spin Hamiltonian via a cluster mean-field approach} \label{subsec:cluster_mean_field}

\begin{figure}
    \centering
    \includegraphics[width=1.0 \linewidth]{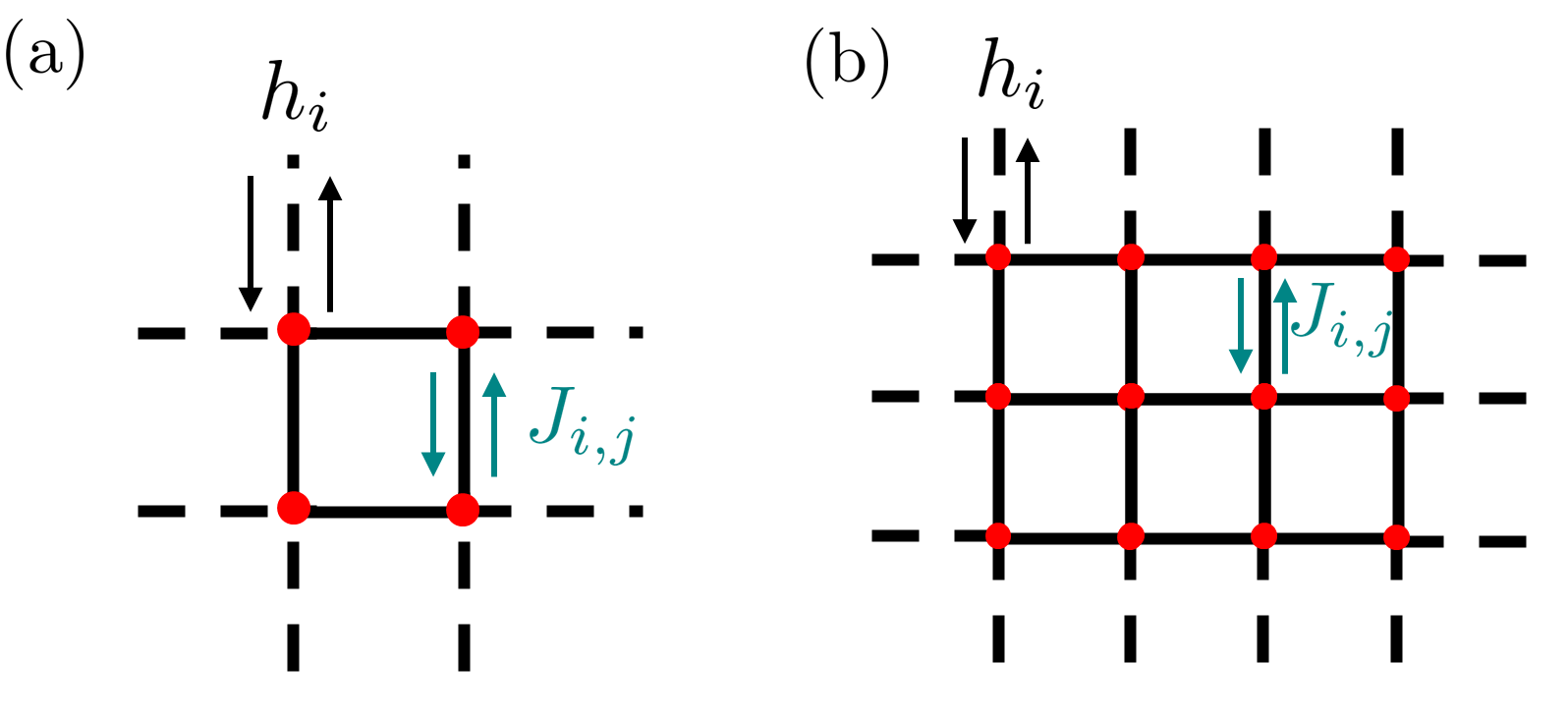}
    \caption{Schematic representation of the cluster geometry for (a) $N=4$ sites and (b) $12$ sites. The dashed black lines represent the interaction with the surrounding mean-field whereas the full black line show the interactions within the cluster. In the twelve sites lattice, the two sites inside the cluster do not interact with the mean-field. }
    \label{fig:CMFA}
\end{figure}

We now focus on the computation of $Q_{ij}$. It requires the computations of the spin-spin correlation function $\langle S_i^z S_j^z \rangle$.
%and hence the pr the ground state $\ket{\psi_0}_\mathrm{S}$ of $H_\mathrm{S}$.

We consider a cluster of $N_\text{x}$ columns and $N_\text{y}$ rows (see Fig.~\ref{fig:CMFA} for an example) surrounded by a mean field. The number of sites in the cluster is then $N = N_\text{x} \times N_\text{y}$.

%and hence the pr the ground state $\ket{\psi_0}_\mathrm{S}$ of $H_\mathrm{S}$.

The cluster mean-field approximation leads  to
\begin{equation}
    S_i^z S_j^z \approx \langle S_i^z \rangle S_j^z + \langle S_j^z \rangle S_i^z - \langle S_i^z \rangle \langle S_j^z \rangle,
\end{equation}
where $i$ ($j$) is inside the cluster at the border of it and $j$ ($i$) is not. The mean-field parameter $\langle S^z_i \rangle$ is the same for all sites in the thermodynamic limit. As we consider finite-size systems,  we numerically compute 
\begin{equation}
    \overline{m} = \frac{1}{N} \sum_{i=1}^{N} \langle S_i^z \rangle.
\end{equation}
This mean magnetization will be the one outside the cluster following a self-consistent loop. 

Therefore, $\sum_{i,j} J_{ij} S^z_{i}S^z_{j} = \sum_{i,j} J_{i,j} (m S_j^z + m S_i^z)  = m \sum_{i,j} J_{i,j} (S_j^z + S_i^z)$, neglecting constant terms. However, the matrix element $J_{i,j}$ is not known for a site $i$ inside the cluster and a site $j$ outside of it. In the thermodynamic limit, all $J_{i,j}$ are equals as it is the one-particle density matrix of a free fermionic system. We can thus take the mean value of all $J_{i,j}$ for nearest neighbors inside the cluster to guess the interaction between sites inside and outside the cluster. Let's then define 
\begin{equation}
\overline{J} = \frac{1}{N_\text{p}} \sum_{\langle i,j \rangle} J_{i,j}
\end{equation}
where the sum goes all over nearest neighbors in the cluster and $N_\text{p}$ is the number of such pairs. In the square lattice, each site has 4 nearest neighbors. We can define a number $z_i$ which is the number of neighbors outside the cluster for site $i$. For example, this number is equal to $0$ for a site which has 4 neighbors inside the cluster.

Finally we obtain a mean-field term

\begin{equation}
    \sum_{i \in \mathcal{C}} h_i S_i^z = 2 \overline{J} \overline{m} \sum_{i \in \mathcal{C}} z_i S_i^z
\end{equation}
and we obtain Eq.~(3) in the main text.

%\begin{figure}
%    \centering
%    \includegraphics[width=0.7\linewidth]{schema/schema_CMFT_iterations.pdf}
%    \caption{Numerical implementation of SSMF combined with CMFT procedure. The initialization is the same as in  The first step is to solve $H_\mathrm{f}(Q)$ to get J. Then, $H_\mathrm{S}^{(C)}(J,h)$ can be solved to get its ground state $\ket{\psi_0}_\mathrm{C}$. At this point, another loop goes on to obtain $h$. With the ground state, we obtain a new $h$ that is used to define a new $H_\mathrm{S}^{(C)}(J,h)$. The CMFT loop goes on until a criteria on $h$ or the maximal number of loop is reached. $Q$ is now calculated from $H_\mathrm{S}^{(C)}(J,h)$ with the last $h$ of the CMFT loop. The SSMF loop continues as described in \ref{SSMF}.}  
%    \label{fig:loop_SSMF_CMFT}
%\end{figure}

%\begin{figure}
%    \centering
%    \includegraphics[width=0.8\linewidth]{schema/schema_SSMF_iterations.pdf}
%    \caption{Numerical implementation of SSMF theory. The first step is to initialize values with a first guess for $Q_0$. Then $H_\mathrm{f}(Q)$ is solved to obtain correlations fermionic correlations. The matrix $J$ is then extracted from these values. $H_\mathrm{S}(J)$ is then diagonalized to obtain spin-spin correlations. The loop goes on until one criteria is reached. The final values of $J$ and $Q$ are the one used to study the physics of the model.}
%    \label{fig:loop_SSMF}
%\end{figure}

\subsection{Convergence of the self-consistent loop}
\begin{figure}
    \centering
    \includegraphics[width=1. \linewidth]{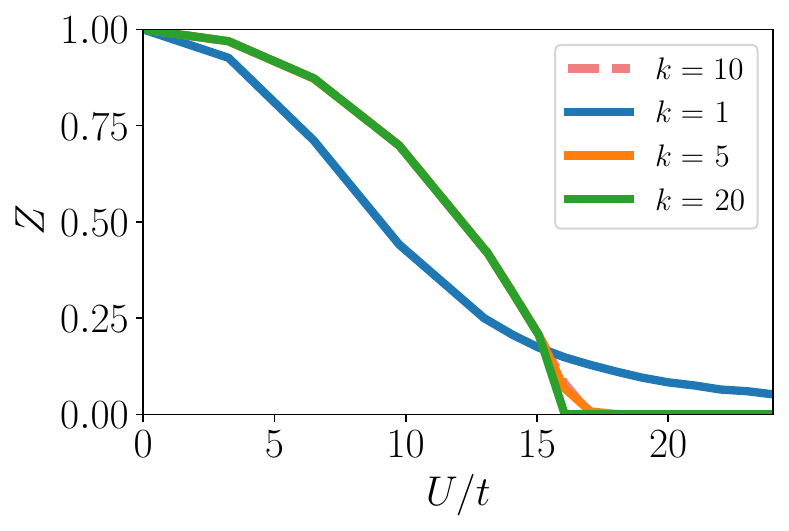}
    \caption{\textit{Impact of imposed number of loops $k$ in the slave-spin mean-field theory for a cluster of $N=4$ sites.} The resolution method is annealing and all sources of error are neglected.
    \label{fig:lmax}
    }
    
\end{figure}

\begin{figure}
    \centering
    \includegraphics[width = 1.0\columnwidth]{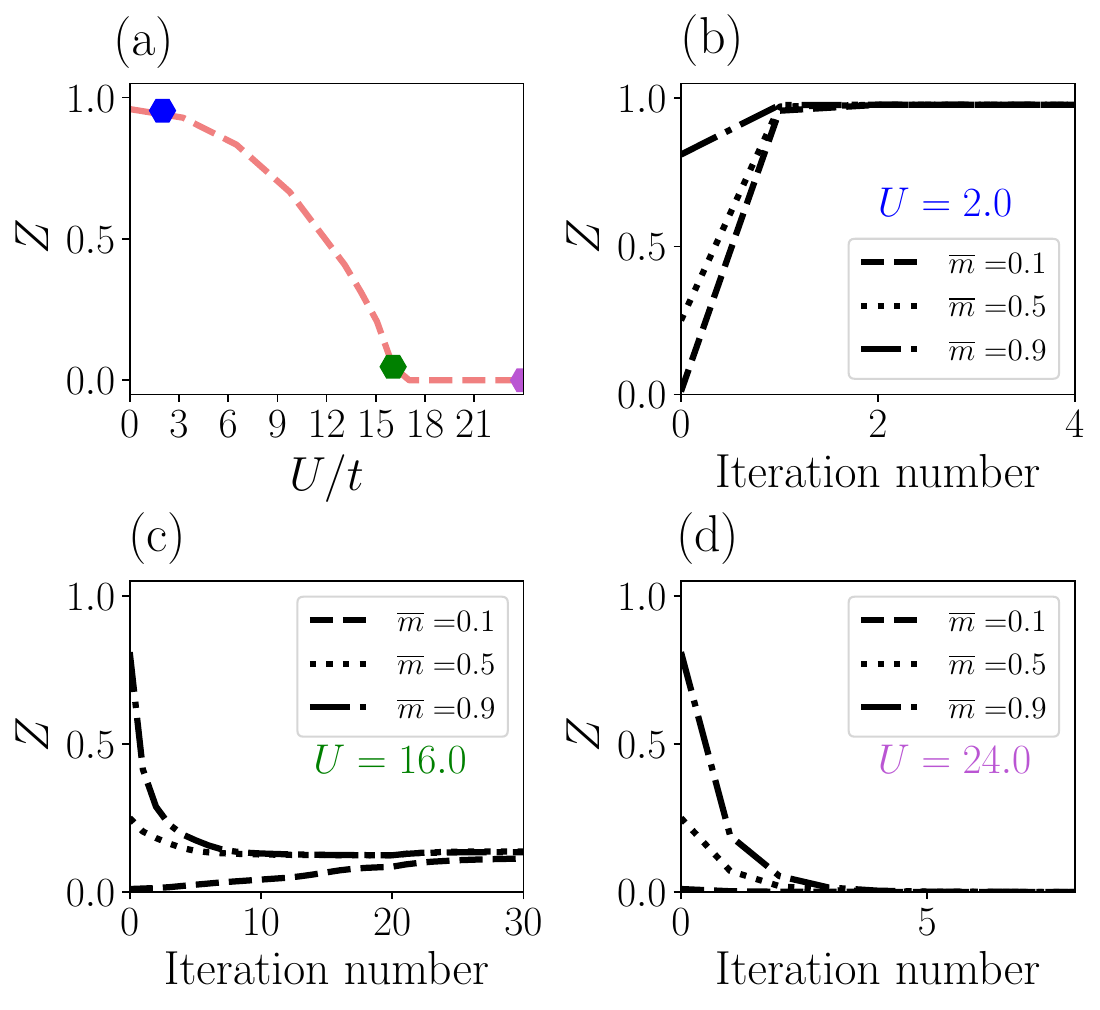}
    \caption{\textit{Evolution of $Z$ as a function of loop iterations for a 6 sites embedding}. (a) Mott transition for a $3 \times 3 $ cluster where three points are highlighted. The convergence of these points ((b) $U/t=2.0$, (b) $U/t=16.0$ and (d) $U/t=24.0$) during the slave-spin mean-field procedure is shown for different initial guess of the mean field $\overline{m}$ (0.1, 0.5 and 0.9). The solving method is annealing where all source of noise are neglected. The number of allowed iteration is increased to $100$ but the x-axis are limited to convergence in the three panels for sake of clarity. and the error accepted is $\eta = 10^{-5}$.}
    \label{fig:h_CMFT_iterations}
\end{figure}

The self-consistent procedure to solve the inner loop is first to guess an initial value for the magnetization $m_0$, then to solve Eq.~(3) in main text and calculate $\overline{m} = \frac{1}{N} \sum_i^{N} \langle S_i^z \rangle$ in the groundstate obtained. The loop goes on until a convergence criteria is reached. In our simulation, two criteria are used: the number of inner loop and outer loop can be narrowed by a number $k$ (so the total number of loop allowed is $k \times k$). The second criteria is the norm of the difference between $Q$ at step $l-1$ and $Q$ at step $l$ for the outer loop and the norm of the difference between $m$ at step $l-1$ and $m$ at step $l$ for the inner loop. We choose a value $\eta$ such as the loop stop if one of the two norm is lower than $\eta$. in our simulation we choose $\eta = 0.01$.
The evolution of $Z$ as a function of iterations is shown in Fig.~\ref{fig:h_CMFT_iterations} for a cluster of $N = 6$ sites, $k=10$ and $\eta = 10^{-6}$. Different initial guess for $\overline{m}$ are tested and they all converge toward the same value which states for the robustness of the method. The convergence takes more time close to the transition value. The impact of the number $k$ imposed is shown in Fig.~\ref{fig:lmax}.

%%%%%%%%%%%%%%%%%%%%%%%%%%%%%%%%%%%%%%%%%%%%%%%%%%%%%%%%%%%%%%%%%%%%%%%%
\section{Solving the spin model with the Rydberg-based processor: details}\label{subsec:rydberg}

In order to solve Eq.~(3) of the main text, we use the Transverse-field Ising Hamiltonian [Eq.~(4) in main text] as naturally implemented on a Rydberg based quantum processor. As discussed in the main text, we apply an annealing procedure with the final Hamiltonian as close as possible to the Hamiltonian whose ground state features the correlation functions we want to compute. In this section, we discuss in more details the annealing procedure and the deviations from the ideal case.

\subsection{Optimization of the geometry} \label{subsec:optim}

The Hamiltonian we are considering, displays a self-consistently determined spin coupling matrix $J_{ij}$, while the Hamiltonian  controlled in the experiment features a van der Waals interaction term $\sum_{j \neq i} C_6/r_{ij}^6$. This section explains how we optimize the geometry of the atom array so that both couplings match as closely as possible.

%Following Eq~(\ref{eq:geometry}), one need to impose the geometry of the Rydberg atoms.
\begin{figure}[!h]
    \centering
    \includegraphics[width = 1 \linewidth]{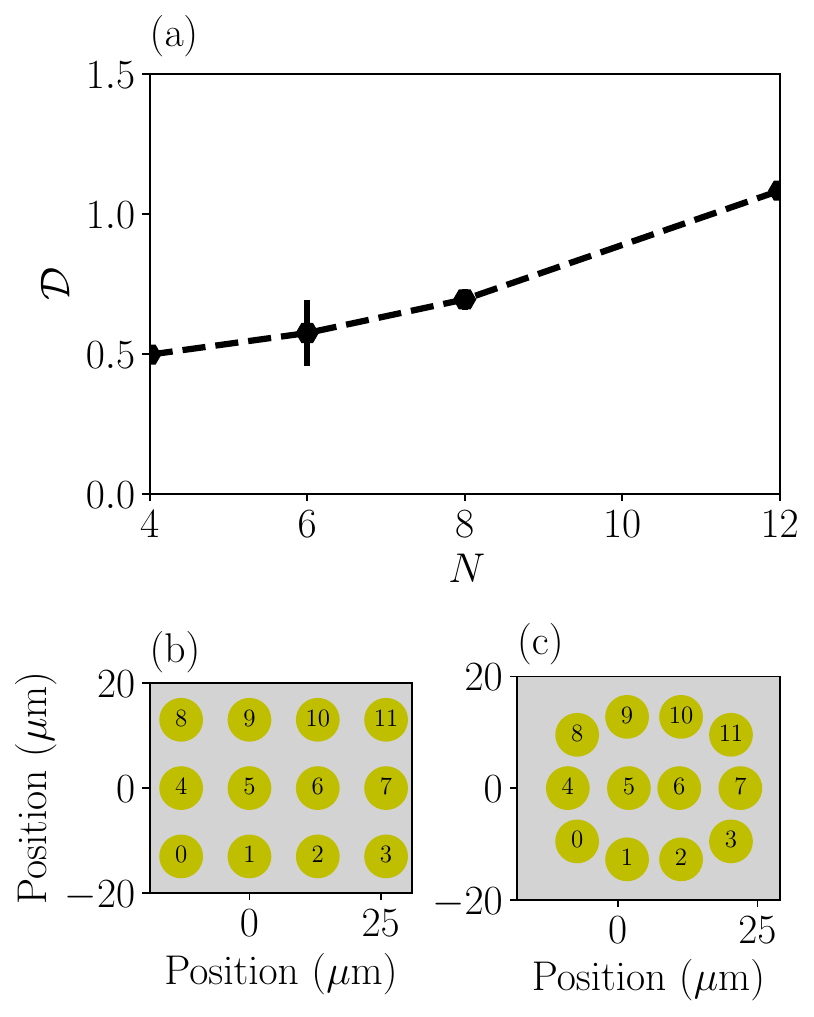}
    \caption{\textit{Optimization of geometry for an implementation on a real device.}(a) Mean value of the cost function $\mathcal{D}$ (Eq.~\ref{eq:norm}) for different cluster size at $U = 13.1 $ MHz. The error bar shows the standard deviation over all $\mathcal{D}$ values encountered during loops. (b) Initial position of the atoms before optimization for a $N=12$ sites cluster in the last outer loop of the slave-spin mean-field method at $U=13.1$ MHz. (c) Position of atoms after the optimization of the geometry to minimize $\mathcal{D}$. }
    \label{fig:D_cost}
\end{figure}

Our goal is thus to minimize the cost function
\begin{equation}\label{eq:norm}
    \mathcal{D} = \sqrt{\sum_{i, j} \left ( \frac{C_6}{|r_{i,j}|^6} + 4J_{i,j} \right )^2 }.
\end{equation}
We use the conjugate gradient descent algorithm from the scipy library \cite{2020SciPy-NMeth}\, with the following initial guess for the geometry: we place the atoms on a square lattice where the distance between nearest-neighbor atoms is $r_\text{init} = \text{max}_{ij}[(C_6/|4J_{ij}|)^{\frac{1}{6}}]$.

The evolution of $\mathcal{D}$ as a function of the number of sites in the cluster is shown Fig.~\ref{fig:D_cost}. The optimization of the positions does not lead to a vanishing $\mathcal{D}$. In practice, the gradient descent algorithm can be trapped in numerous local minima, leading to a poor approximation of $-4\times J_{ij}$ by the interaction matrix element. In addition, difficulties can arise directly from the symmetries of the initial cluster guess. For instance, in the case of a $2 \times 2$ cluster with nearest-neighbor distance $a$, the distance between next nearest neighbors is always $a/\sqrt{2}$ whereas it should be $0$ for our model since $J_{ij} = 0$ for next nearest neighbors. Therefore, in most cases, $\mathcal{D}$ is not exactly zero and finding the best geometry is not an easy task.

\begin{figure}
    \centering
    \includegraphics[width=1. \linewidth]{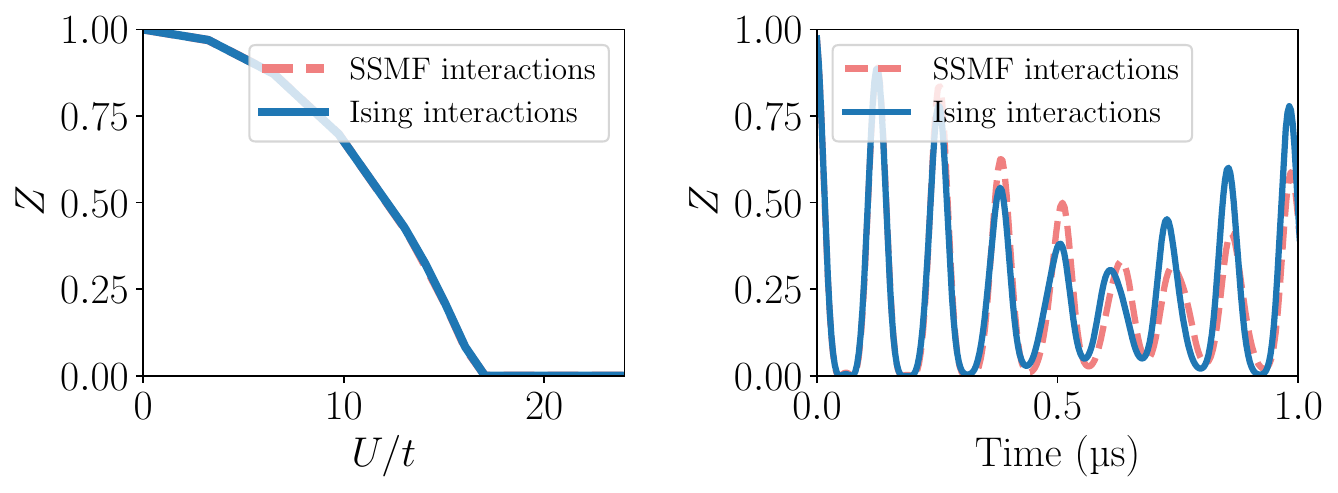}
    \caption{\textit{Impact of considering a realistic geometry on a cluster of $N=4$ sites}. \textit{Left:} Comparison of $Z$ values between method with the real matrix $J$ and the optimized one for 4 sites. \textit{Right:} $Z$ dynamics after a quench $U_\mathrm{f} = 13$ MHz with the same comparison. All other sources of noise are neglected. }
    \label{fig:optimized_geo}
\end{figure}

Despite these problem, we find that the impact of considering an imperfect optimization of the geometry (leading to a nonzero $\mathcal{D}$) does not lead to significant changes. In Fig.~\ref{fig:optimized_geo}, we show the outcome of the equilibrium and out-of-equilibrium computations with a "perfect geometry" (assuming the coupling is actually $J_{ij}$) and an imperfect geometry.
For the Mott transition, we observe negligible differences for $N=4$ cluster. 
For the dynamical behavior, we do observe a change in amplitude  but the frequency remains the same as for the slave-spin mean-field interactions.
To illustrate the outcome of the optimization procedure, we also show an example of initial and optimized position for $N = 12$ cluster sites in Fig.~\ref{fig:D_cost}.
There, the final pattern is slightly distorted compared to the translation-invariant initial pattern. This is due to the fact that the couplings at the edges of the cluster differ from the ones in the "bulk" of the cluster to account for the cluster's environment. As the cluster size grows, these edge effects have decreasing influence, and the optimization becomes easier. We thus conclude that the geometry optimization yields reasonably faithful interactions. 

%Indeed, the optimization of the geometry is far from being the main source of errors in our noisy simulations. Symmetries in the matrix $J$ can in fact help to make the optimization easier and at the thermodynamic limit, $J$ should be homogeneous at it contains two-body correlators of a free fermionic system. 

%We use the exact gradient... \ta{give formula used for gradient}\am{C'est-à-dire ? La formule exacte de ce qu'il se passe dans la function de Scipy ? C'est vraiment nécessaire tu crois si je précise que c'est le conjugate gradient algorithm ?} \ta{non. Tu ne donnes pas à la fonction de scipy f(x) et df/dx ? tu donnes seulement f(x) ? je croyais que tu donnais aussi df/dx (avec f= la fonction de coût en C1). Par ailleurs, quelle option donnes-tu à scipy.minimize ? quelle méthode ? "CG" ?}\am{J'utilise effectivement "CG" et non je donne pas la dérivée}

%\ta{ici ce serait important de dire un mot des résultats quantitatifs de cette procédure: est-ce que les positions trouvées permettent de reproduire le bon $J_{ij}$ ? comment cela dépend-il de la taille du cluster? Comme tu le sais de ton papier précédent, ce n'est pas du tout évident qu'il existe une solution à ce problème (il n'y a pas toujours des positions qui permettent d'atteindre un J donné). Par exemple, tu pourrais donner le $\mathcal{D}$ obtenu pour un $U$ fixé pour différentes tailles (tu dois avoir cette info quelque part). Aussi, il serait instructif de montrer un plot des positions initiales vs finales...}\am{J'ai fait un petit paragraphe pour discuter de cela et j'ai rajouter les plots.}

\subsection{Details of the annealing schedule}

Once the geometry is found, the atoms are prepared in the state $\ket{\psi_\mathrm{start}} = \ket{g}^{\bigotimes N}$.
The following Hamiltonian is the one applied at $\tau=0$
\begin{equation}\label{eq:Hstart}
    H_\mathrm{start} = \sum_{i\ne j}\frac{C_6}{|\textbf{r}_i-\textbf{r}_j|^{6}} \hat{n}_i \hat{n}_j -\hbar \delta_\mathrm{start} \sum_i n_i
\end{equation}
where $\delta_\mathrm{start}/(2\pi)=- 5$ MHz so that $\ket{\psi_\text{start}}$ is the most excited state of (\ref{eq:Hstart}).
The Rabi frequency and the detuning, addressing globally the array, are then varied over duration $\tau_\text{max}$ to reach the Hamiltonian $-H_\mathrm{s}^{\mathcal{C}}$.
Following the procedure described in the main text, the Rabi frequency starts at $0$ MHz and is driven linearly to $U/2$ ($\hbar\Omega(\tau_\mathrm{max}) = U/2$). Similarly, the detunings are all prepared at a value $\delta_\mathrm{start}$ and are driven separately to values $\hbar\delta_i(\tau_\mathrm{max}) = \sum_{j \neq i} C_6/r_{ij}^6 + 4\overline{J}\overline{m}z_i$.

%Observables ($Q$, $Z$, $g$) are measured with a sample of shots obtained from the state of the atoms.

\begin{figure}
    \centering
    \includegraphics[width=1. \linewidth]{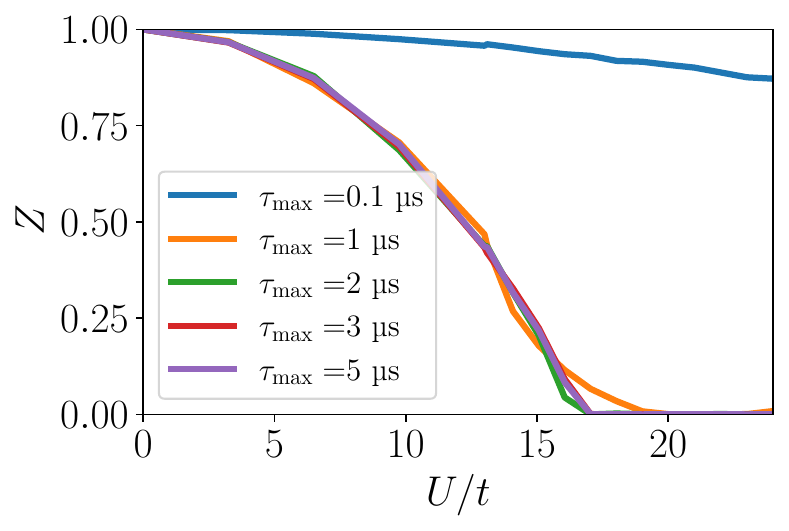}
    \caption{\textit{Impact of total annealing time for a cluster of $N = 4$ sites.} All other sources of error are neglected. 
    \label{fig:plot_sev_tmax}
    }  
\end{figure}

The effect of the annealing time on the outcome of the simulation is shown in Fig.~\ref{fig:plot_sev_tmax}. Starting from $\tau_\text{max} = 3 \, \, \mu$s, its influence is negligible. In our simulations, we choose  $\tau_\text{max} = 4 \, \mu$s to ensure a good convergence.

\begin{figure}
    \centering
    \includegraphics[width=1. \linewidth]{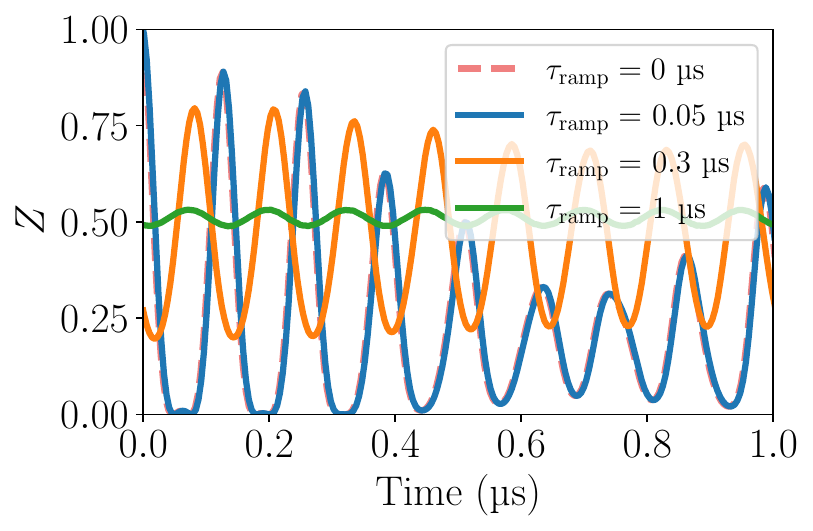}
    \caption{\textit{Impact of switch-on time $\tau_\mathrm{ramp}$ in the quench dynamic for a cluster of $N = 4$ sites.} We consider $U_\mathrm{f} = 13$ MHz. All other sources of error are neglected. 
    \label{fig:tramp}
    }
    
\end{figure}

To study  the dynamics in the Hubbard model, we need to quench the value of the Rabi term. In practice, the quench is not instantaneous owing to the finite response time of the optical modulators. In Fig.~\ref{fig:tramp}, we investigate the effect of the finite switch-on time $\tau_\mathrm{ramp}$ on the Rabi frequency and the detuning. We see that this time does impact the frequency of the signal for $\tau_\mathrm{ramp} \geq 0.3 \, \mu$s. For the modulator technical specifications $\tau_\mathrm{ramp} \approx 0.05\, \mu$s, and the effects are negligible.  

\subsection{Experimental imperfections}

The algorithm described in the main text is designed to work on existing Rydberg processors. To  evaluate the effects of noise and experimental limitations on the results, we include them in the simulation. Here we explain the methods we use to emulate the noise and how they are implemented in our code.
All numerical simulations are performed with the library QuTiP \cite{johansson_qutip_2013} (exact diagonalization) and the Quantum Learning Machine. The SPAM error is implemented with the library Pulser \cite{silverio_pasqal-iopulser_2022}.

\subsubsection{Dephasing noise}
\begin{figure}
    \centering
    \includegraphics[width=1.\linewidth]{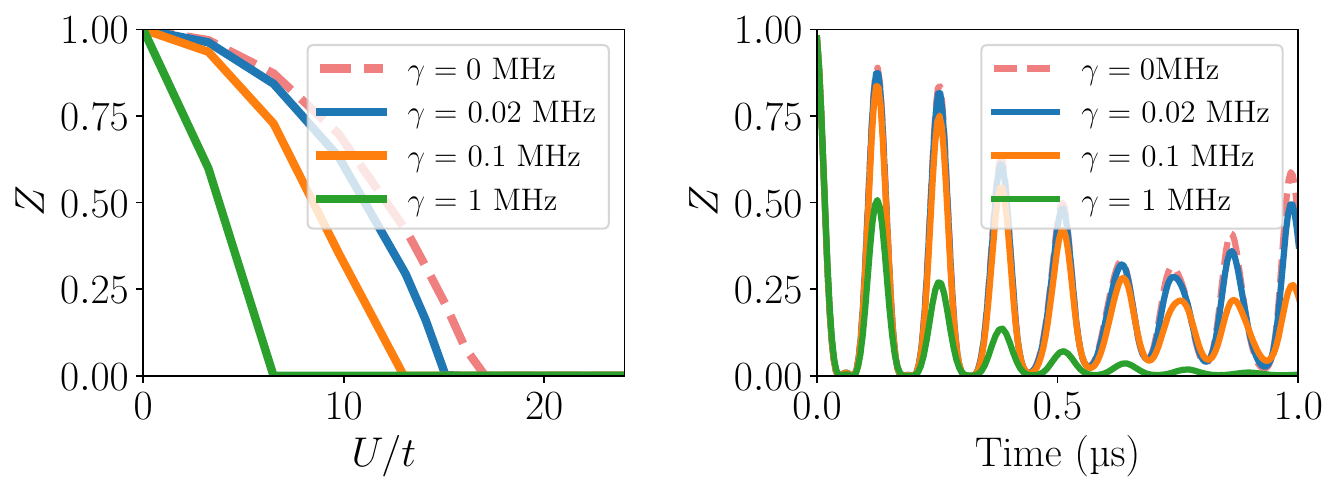}
    \caption{\textit{Effect of dephasing noise on the result \sout{of} for a cluster of $N = 4$ sites.} \textit{Left:} Impact one the Mott phase transition. All other sources of noise are neglected. \textit{Right:} Impact on the quench dynamic for $U_\mathrm{f} = 13$ MHz.  }
    \label{fig:gamma}
\end{figure}

Decoherence during the annealing procedure is described via the Lindblad master equation (following \cite{lienhard_observing_2018}):
\begin{equation}
    \frac{d \rho}{d\tau} = -i[H(\tau),\rho] - \frac{1}{2}\sum_{i =1}^{N} \gamma_i \Big [ \Big \{ L_i^{\dagger} L_i, \rho \Big \} -2L_i \rho L_i^{\dagger} \Big ] 
    \end{equation}
where $\rho$ is the density matrix of the system and $H(\tau)$ is the resource Hamiltonian at a time $\tau$ during the annealing.
The jump operators $L_i$ corresponding to dephasing are equal to $n_i$. 
For the sake of simplicity, we use a single dephasing parameter $\gamma_i = \gamma$. The effect  of this noise is shown for various parameters $\gamma$ in Fig.~\ref{fig:gamma}. For the quench dynamics, the dephasing damps the oscillations but does not change the frequency. We can see this behaviour on the position of the Mott transition, which is towards small values of $U/t$ for large $\gamma$.

\subsubsection{Sampling and measurement error}

\begin{figure}
    \centering
    \includegraphics[width=1.\linewidth]{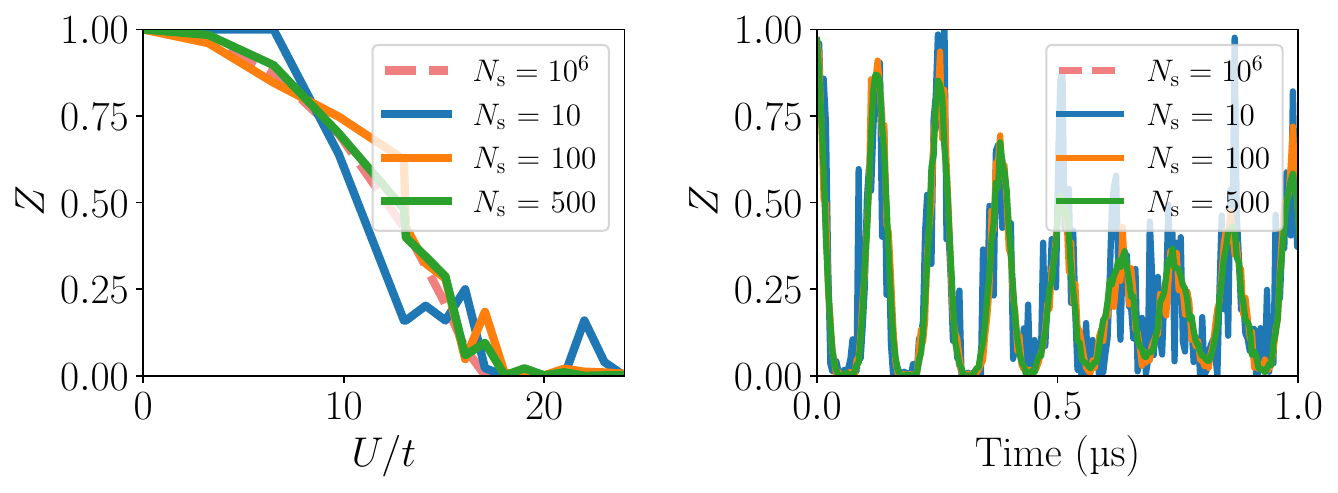}
    \caption{\textit{Impact of different sampling rate $N_\mathrm{s}$ on measured states for a cluster of $N = 4$ sites.} \textit{Left:} at equilibrium and \textit{right:} out of equilibrium ($U_\mathrm{f} = 13$ MHz). All other sources of error are neglected. }
    \label{fig:shotnoise}
\end{figure}

\begin{figure}
    \centering
    \includegraphics[width=1. \linewidth]{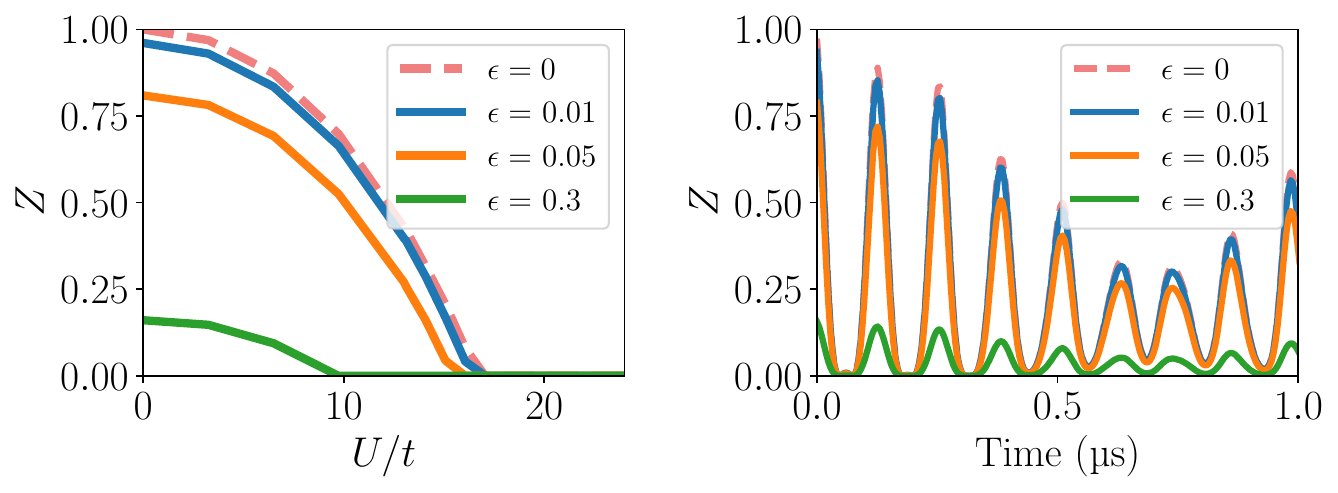}
    \caption{\textit{Impact of measurement error for a cluster of $N = 4$ sites}. \textit{Left:} shows the effect of $\epsilon = \epsilon'$ at equilibrium and \textit{right:} out of equilibrium for $U_\mathrm{f} = 13$ MHz. The number of shots considered for each measurement is $10^{6}$. All other sources of noise are neglected. }
    \label{fig:eps}
\end{figure}

We simulate the sampling of states as would be done on Rydberg processor (shot-noise) by picking randomly $N_\text{s}$ times a bitstring with a probability equal to the probability to measure this bitstring in the z-basis. The impact of such a procedure is shown Fig.~\ref{fig:shotnoise} at equilibrium and out of equilibrium: it becomes negligible as soon as $N_\mathrm{s} \approx 100$. 

Finally, we model the readout error by a probability of error $\epsilon$ of detecting an atom in a state $\ket{r}$ instead of its real state $\ket{g}$ and $\epsilon'$ of not detecting an excited atom. We choose  $\epsilon = \epsilon' = 3\%$ for both values in the main text \cite{de_leseleuc_analysis_2018}. The impact of the errors for $\epsilon = \epsilon'$ is shown in Fig.~\ref{fig:eps}. Until $\epsilon \approx 5 \%$, the behavior of the system remains the same.

\bibliography{main}

\end{document}